\documentclass[%
 preprint,
superscriptaddress,
 amsmath,amssymb,
 aps,
 prl,
 letterpaper,
 raggedbottom,
 nobalancelastpage]{revtex4-1}

\usepackage[english]{babel}
\usepackage[utf8x]{inputenc}
\usepackage[T1]{fontenc}

\usepackage[a4paper,top=3cm,bottom=2cm,left=2cm,right=2cm,marginparwidth=1.75cm]{geometry}

\usepackage{amsmath}
\usepackage{units}
\usepackage{graphicx}
\usepackage[colorinlistoftodos]{todonotes}
\usepackage[colorlinks=true, allcolors=blue]{hyperref}
\usepackage{setspace}
\usepackage{lineno}
\usepackage{graphicx}
\usepackage{dcolumn}
\usepackage{bm}

\begin{document}
\title{Experimental observation of plasma wakefield growth driven by the seeded self-modulation of a proton bunch}

\author{M.~Turner\textsuperscript{*}}
\affiliation{CERN, Geneva, Switzerland}
\author{E.~Adli}
\affiliation{University of Oslo, Oslo, Norway}
\author{A.~Ahuja}
\affiliation{CERN, Geneva, Switzerland}
\author{O.~Apsimon}
\affiliation{University of Manchester, Manchester, UK}
\affiliation{Cockcroft Institute, Daresbury, UK}
\author{R.~Apsimon}
\affiliation{Lancaster University, Lancaster, UK}
\affiliation{Cockcroft Institute, Daresbury, UK}
\author{A.-M.~Bachmann}
\affiliation{CERN, Geneva, Switzerland}
\affiliation{Max Planck Institute for Physics, Munich, Germany}
\affiliation{Technical University Munich, Munich, Germany}
\author{M.~Barros Marin}
\affiliation{CERN, Geneva, Switzerland}
\author{D.~Barrientos}
\affiliation{CERN, Geneva, Switzerland}
\author{F.~Batsch}
\affiliation{CERN, Geneva, Switzerland}
\affiliation{Max Planck Institute for Physics, Munich, Germany}
\affiliation{Technical University Munich, Munich, Germany}
\author{J.~Batkiewicz}
\affiliation{CERN, Geneva, Switzerland}
% ...
\collaboration{The AWAKE Collaboration}
\noaffiliation

\sloppy

\maketitle

\textbf{We measure the effects of transverse wakefields driven by a relativistic proton bunch in plasma with densities of $2.1\times10^{14}$ and \unit[$7.7\times10^{14}$]{electrons/cm$^3$}. We show that these wakefields periodically defocus the proton bunch itself, consistently with the development of the seeded self-modulation process. We show that the defocusing increases both along the bunch and along the plasma by using time resolved and time-integrated measurements of the proton bunch transverse distribution. We evaluate the transverse wakefield amplitudes and show that they exceed their seed value (\unit[$<$15]{MV/m}) and reach over \unit[300]{MV/m}. All these results confirm the development of the seeded self-modulation process, a necessary condition for external injection of low energy and acceleration of electrons to multi-GeV energy levels.}

Particle-driven plasma wakefield acceleration offers the possibility to accelerate charged particles with average accelerating gradients of the order of GV/m over meter-scale distances \cite{CHEN,BLUMENFELD}. The distance over which plasma wakefields can be sustained depends, among other parameters, on the energy stored in the relativistic drive bunch. It was demonstrated that a \unit[42]{GeV} electron bunch can increase the energy of some trailing electrons by \unit[42]{GeV} over a distance of \unit[0.85]{m} \cite{BLUMENFELD}. Reaching much higher witness bunch energies would require staging of multiple acceleration stages, each excited by a new drive bunch. Staging is however experimentally challenging \cite{CLindstrom}. Using a proton bunch to drive wakefields can help overcome the need for staging since available proton bunches, for example at CERN, carry enough energy to drive GV/m plasma wakefields over hundreds of meters in a single plasma \cite{PROTONDRIVEN,PoP18-103101}. 

The maximum accelerating field depends on the plasma electron density $n_{pe}$ and can be estimated from the cold plasma wave-breaking field \cite{DAWNSON} $E_{max}=m_e \omega_{pe} c/e$, where $m_e$ is the electron mass, $\omega_{pe} = \sqrt{n_e e^2/\epsilon_0 m_e}$ is the angular electron plasma frequency, $c$ is the speed of light, $e$ is the electron charge and $\epsilon_0$ is the vacuum permittivity. To reach GV/m fields, the plasma electron density has to exceed \unit[$10^{14}$]{cm$^{-3}$}. At these densities, the plasma wavelength $\lambda_{pe}= 2 \pi c/\omega_{pe}$ is shorter than \unit[3]{mm}. 

From linear theory, the root-mean-square (rms) drive bunch length $\sigma_z$ optimal to drive wakefields for a given plasma density is on the order of the plasma wavelength $\lambda_{pe}$ and can be expressed as $k_{pe}\sigma_z \cong \sqrt{2}$, where $k_{pe}=\omega_{pe}/c$ \cite{LINEARTHEORY}. The shortest high-energy proton bunches available have an rms length of \unit[$\sigma_{z}=$6-12]{cm}. When satisfying $k_{pe}\sigma_z \cong \sqrt{2}$, these proton bunches are therefore much too long to drive GV/m wakefield amplitudes. However, when the bunch is much longer than the plasma wavelength it is subjected to a transverse instability called the self-modulation instability (SMI) \cite{PoP2-1326,PoP4-1154,PPCF53-014003} or, when seeded, the seeded self-modulation (SSM) \cite{MUGGLI,EPAC98-806}.

When a long proton bunch enters the plasma, it drives transverse and longitudinal wakefields with a period determined by the plasma electron density ($\lambda_{pe} \propto \sqrt{n_{pe}}$) and an amplitude also determined by the drive bunch parameters \cite{PoP21-123116-on-axis}. The transverse wakefields are periodically focusing and defocusing and act back on the bunch itself. Where transverse fields are defocusing, the bunch radius increases and the bunch density $n_b$ decreases. Where they are focusing, the bunch radius decreases, creating regions of higher bunch density that drive stronger wakefields ($\propto n_b$) and thus create the feedback loop for the self-modulation process \cite{PoP22-103110-smisim}. 

The regions of focused protons are spaced by $\lambda_{pe}$ and form a train of microbunches. Each microbunch satisfies $k_{pe}\sigma_z \cong \sqrt{2}$ and the bunch train can thus resonantly drive large amplitude plasma wakefields. During the self-modulation process the wakefield amplitude grows both along the bunch and along the plasma. Seeding ensures that: a) the bunch self-modulates; b) the phase of the wakefield is stable and reproducible \cite{PoP21-083107,NAPAC16-684}; and c) that the hose instability \cite{HOSING} (with a comparable growth rate) is suppressed.

The successful and controlled development of the SSM is a necessary requirement to be able to use long proton bunches to drive large amplitude wakefields and to accelerate particles ($e^+,e^-$) in these wakefields. Previous work showed self-modulation resulting in the formation of two \cite{Gross} or a few microbunches \cite{SMIRESULTS}. In \cite{Gross} the authors claim that the instability grew above seed level, but the argument is based on simulation results. %
Results of the Advanced Wakefield Experiment (AWAKE) show formation of a large number of proton microbunches (up to 100). These results also show agreement between the measured modulation frequency and the plasma frequency over an order of magnitude in plasma densities \cite{karl}. %

In this \textit{Letter}, we demonstrate that a highly relativistic proton bunch self-modulates radially. We show unambiguous experimental proof of wakefield growth along the plasma and along the bunch. We conclude that as a result of the growth, the driven wakefield amplitudes reached hundreds of MV/m which is much larger than the initial seed level.

In AWAKE \cite{MUGGLI, APATHTO, EDDA}, and for the measurements presented in this \textit{Letter}, we used the following proton bunch parameters: a population of \unit[$(0.5-3)\times10^{11}$]{particles/bunch}, an rms length \unit[$\sigma_z = 6-8$]{cm}, a radial size at the plasma entrance  \unit[$\sigma_r=\sim$0.2]{mm} and a normalized bunch emittance \unit[$\epsilon_N=$2.2]{mm$\cdot$mrad}. 

A \unit[10]{m}-long vapor source \cite{PLASMA1,PLASMA2} provides a rubidium density adjustable in the \unit[$n = (1-10)\times10^{14}$]{atoms/cm$^3$} range. A \unit[120]{fs}, \unit[$<$450]{mJ} laser pulse ionizes the outermost electron of each rubidium atom creating a plasma with a radius of approximately \unit[1]{mm} \cite{MUGGLI}. The laser pulse creates a relativistic ionization front, much shorter than the wakefield period, that effectively seeds the wakefields \cite{VIERA}. When the ionization front is placed near the middle of the proton bunch, the seed wakefields reach an amplitude of a few MV/m, far above the expected noise amplitude of a few tens of kV/m \cite{LOTOVSEEDING}. From this initial seed amplitude, the wakefields and the proton bunch modulation grow along the bunch and plasma.

\begin{figure}[htb!]
\centering
		\includegraphics[width = 0.7\columnwidth]{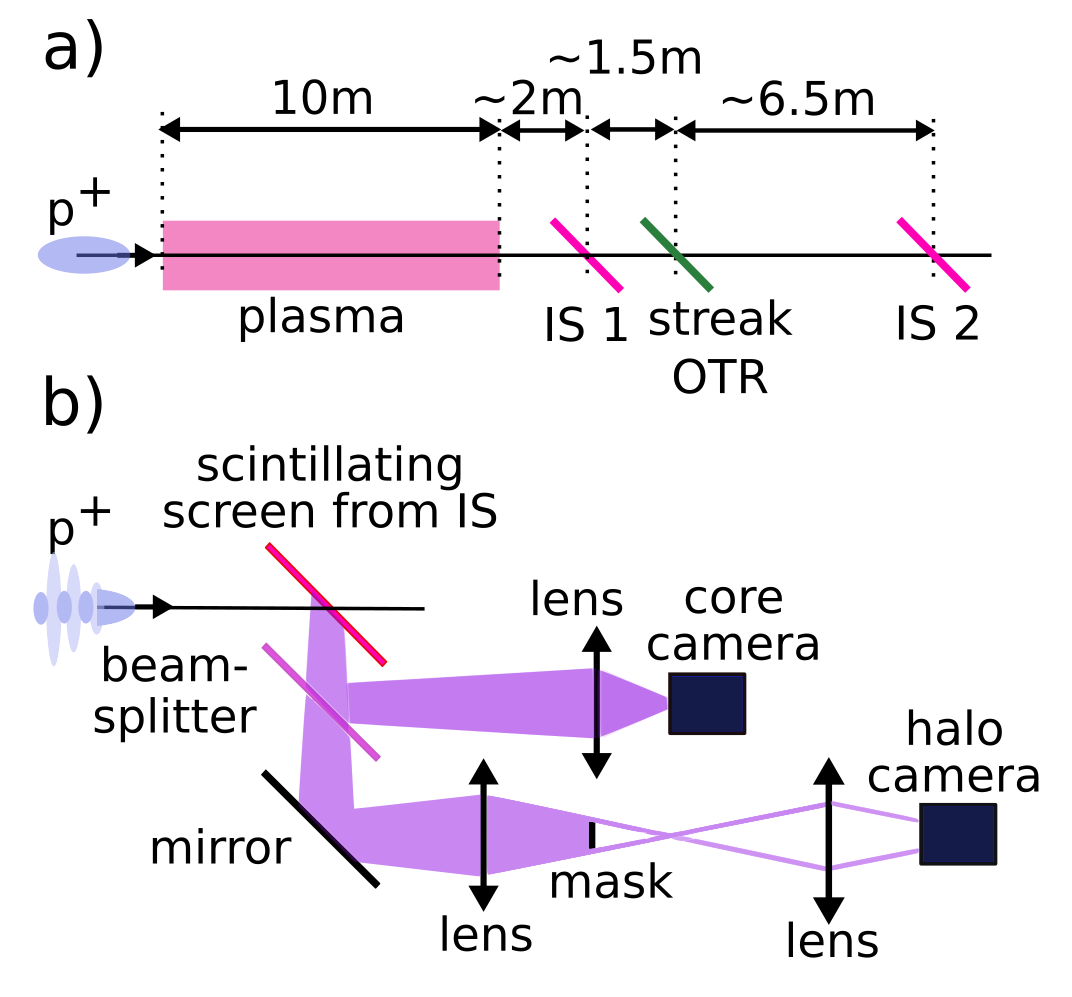}
		\caption{a) Schematic location of the imaging stations (IS1 and IS2) and of the OTR streak camera screen with respect to the plasma. The proton bunch moves from left to right. b) Schematic drawing of the optical setup of the imaging stations.}
		\label{fig:setupofIS}
\end{figure}

As shown in Fig. \ref{fig:setupofIS}a), to experimentally diagnose proton bunch self-modulation, we measure:
a): the structure of the bunch in space and time with a streak camera \cite{karl,STREAK} and the time-integrated transverse distribution with imaging stations (IS) \cite{TurnerIPAC2017}.

The streak camera produces an image of the transverse bunch distribution as a function of time, with pico-second resolution. As protons traverse an aluminium coated silicon wafer, they emit forward and backward optical transition radiation (OTR). 
The backward OTR is imaged onto the entrance slit of the streak camera.  

\begin{figure}[htb!]
\centering
		\includegraphics[width =\columnwidth]{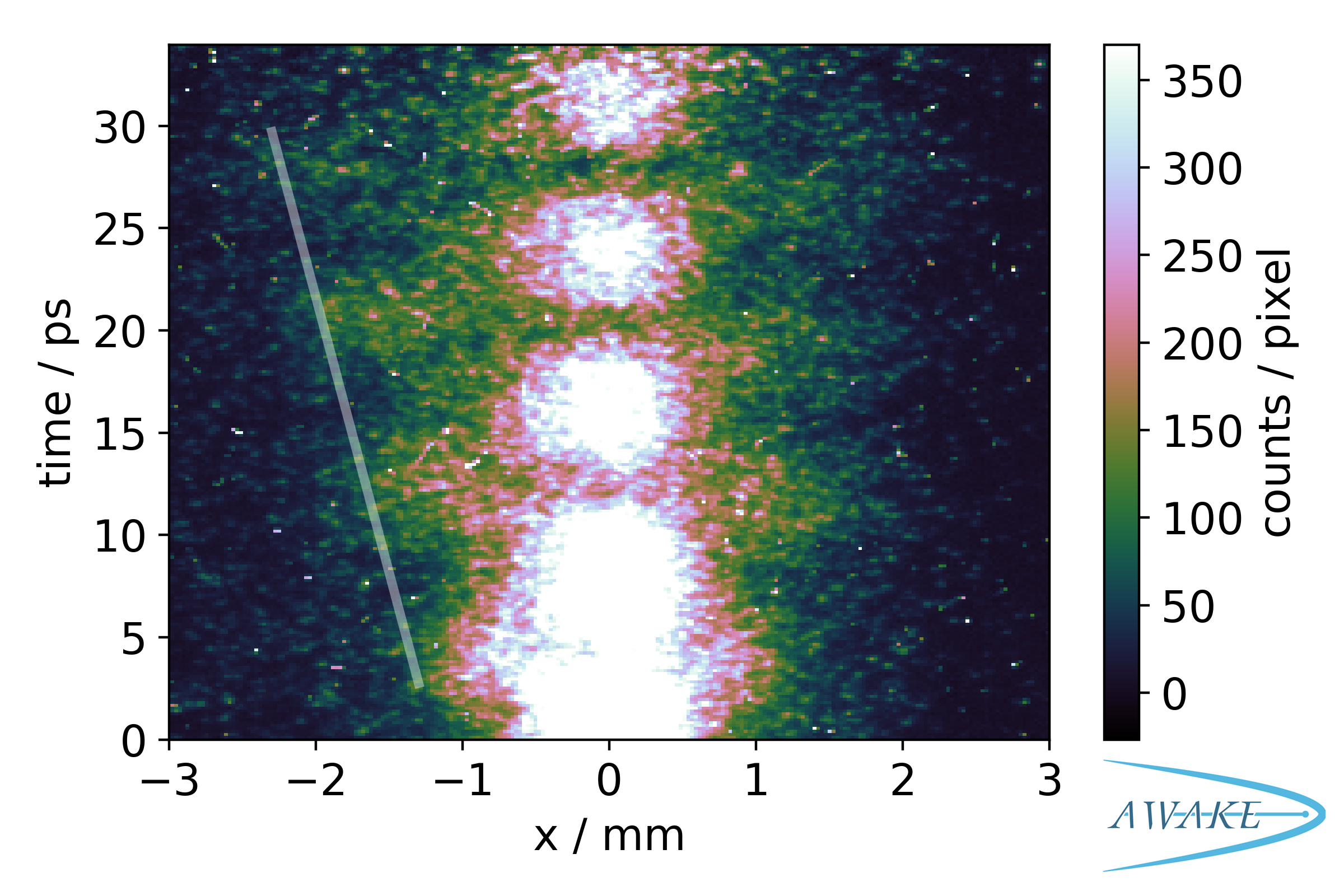}
		\caption{Streak camera image showing the transverse distribution of the self-modulated proton bunch as a function of time. The image is obtained by summing ten individual measurements. 
The bunch moves down along the time axis. The time scale is set to show only \unit[34]{ps} of the \unit[$\sim$73]{ps} image. The white line indicate the observed increase of maximum defocusing of the protons along the bunch.}
		\label{fig:streakimage}
\end{figure}

Figure \ref{fig:streakimage} shows a streak camera image of the first few modulation periods of the proton bunch for a plasma density of \unit[2.1$\times 10^{14}$]{electrons/cm$^3$}. We observe regions of higher and lower light intensity along the time axis, corresponding to  higher and lower proton densities. Regions of focused protons are observed at times $\sim16,\sim24$ and \unit[$\sim32$]{ps}, defocused protons are observed in between  at times: $\sim12,\sim20$ and \unit[$\sim28$]{ps}. The image clearly shows that the maximum transverse position at which protons are observed increases along the bunch (\unit[1.5]{mm} at around \unit[2]{ps}, to \unit[2.5]{mm} at around \unit[30]{ps}), as indicated by the white line in  Fig. \ref{fig:streakimage}. At later times the defocused proton density falls below the detection threshold of the streak camera.


For our proton bunch with $\sigma_z \gg \lambda_{pe}$ and seeded at the peak, the initial transverse wakefields near the entrance of the plasma and the seed point are either zero or focusing and their maximum amplitude is essentially constant (or decreasing)  over the first wakefield periods. Figure \ref{fig:streakimage} shows periodic zones of focused and defocused protons. This indicates that the wakefields developed to include defocusing fields and that their amplitude increases along the bunch. This clearly demonstrates growth of the self-modulation along the bunch. On the image, the effect appears to be slightly asymmetric as the light transport optics setup with limited aperture clips the light on the right-hand side of the image. 

Since wakefields driven by a train of microbunches increase along the train, protons are defocused to much larger radii further along the bunch. Measurements at the first imaging station (see Fig. \ref{fig:setupofIS}a)), located \unit[1.5]{m} upstream of the the streak camera screen, show that the maximum radius of the defocused protons reaches \unit[$\sim7$]{mm} in radius, much larger than the \unit[$\sim$2-3]{mm} visible on Fig. \ref{fig:streakimage}.

To overcome the dynamic range limitations of the streak camera and to detect the most defocused protons, we measured the transverse, time-integrated proton bunch charge distribution with two imaging stations (IS) installed \unit[$\sim$2]{m} (IS 1) and \unit[$\sim$10]{m} (IS 2) after the plasma exit (see Fig. \ref{fig:setupofIS}a).  The IS consist of a scintillating Chromox (Al$_2$O$_3$:Cr$_2$O$_3$) screen mounted inside a stainless-steel vacuum vessel. A schematic drawing of the setup of an IS is shown in Fig. \ref{fig:setupofIS}b). 

The light output of the scintillator is proportional to the energy deposited by the protons in the screen material. Since the energy of all protons remains within \unit[$\pm10$]{GeV} of their initial \unit[$\sim$400]{GeV}, we take the light intensity to be proportional to the number of protons. The emitted light is imaged onto a digital camera. 

In order to record at the same time the proton bunch core (\unit[$\sim10^9$]{protons/mm$^2$}) and the defocused protons (\unit[$\sim10^6$]{protons/mm$^2$}), we split the emitted light with a beam-splitter and send it to two cameras: the 'core camera' records the entire charge distribution; for the halo camera, we block the light emitted by the bunch core with a mask. The mask is placed in the image plane of the first lens imaging the Chromox screen and is re-imaged onto the camera by the second lens. 

\begin{figure}[htb!]
\centering
		\includegraphics[width = 0.7\columnwidth]{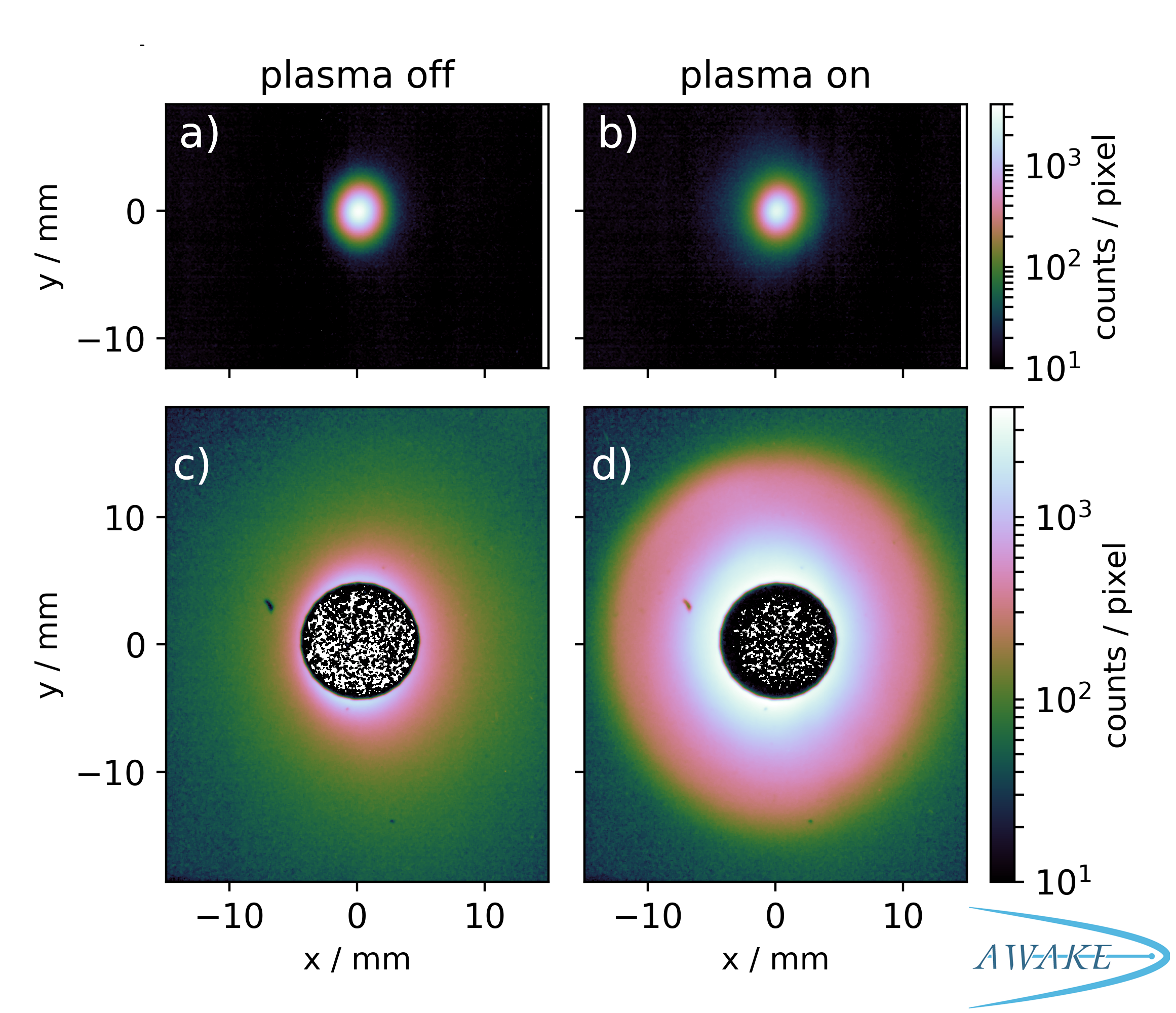}\\
        \includegraphics[width = 0.55\columnwidth]{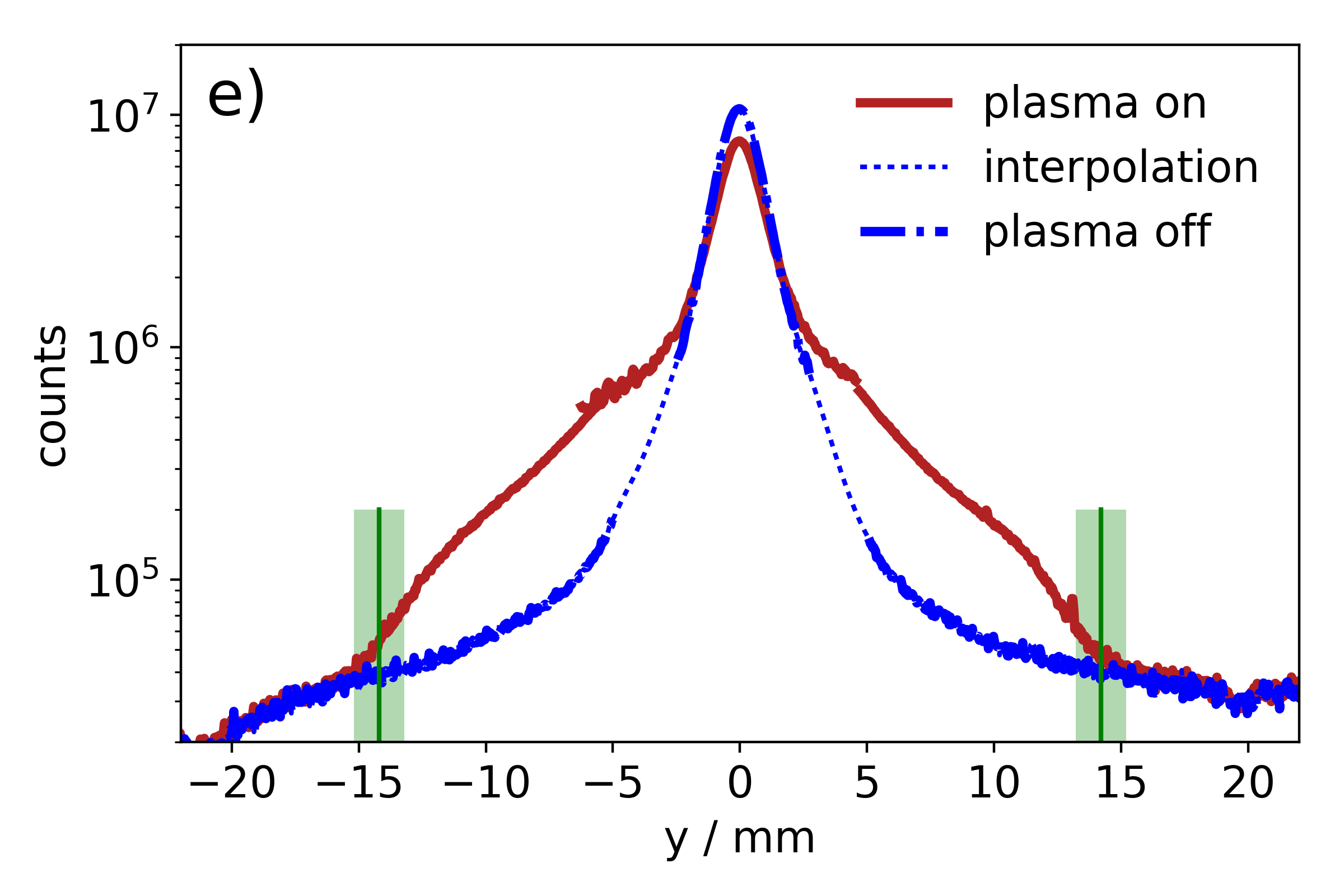}
		\caption{Time-integrated proton bunch charge distributions after the bunch propagated in rubidium vapor with a density of \unit[$7.7\times10^{14}$]{atoms/cm$^3$} (a,c) and plasma with a density of \unit[$7.7\times10^{14}$]{electrons/cm$^{3}$} (b,d). Panels a) and b) show core camera and panels c) and d) halo camera measurements. In panels c) and d), the center of the proton bunch is blocked by a mask. Note the logarithmic scales. e) Combined projections of the images of Fig. \ref{fig:plasmaonoff}a-d). The blue lines show the measurement without plasma and the red lines with plasma. Green vertical bars show the maximum radius of the proton distribution, defocused by the plasma.}
\label{fig:plasmaonoff}
\label{fig:maxradius}
\end{figure}

We show two different measurements at IS 2 (bunch parameters as stated above). Figures \ref{fig:plasmaonoff}a) and 
b) show the core camera images and Figs. 
c) and 
d) show the same events as measured by the halo camera.  On Figs. \ref{fig:plasmaonoff}a) and 
c), we show the proton bunch after propagation in \unit[10]{m} of rubidium vapor at a density of \unit[$7.7\times10^{14}$]{atoms/cm$^3$} (inferred from measurements of the rubidium density \cite{ERDEM}), with no ionizing laser pulse, i.e. no plasma. The images show the transverse distribution of the unmodulated bunch.

Figures \ref{fig:plasmaonoff}b) and 
d) show the proton distribution after propagation in \unit[10]{m} of plasma. The ionizing laser pulse co-propagated at the center of the proton bunch creating a plasma with a density of \unit[$7.7\times10^{14}$]{electrons/cm$^3$}. Note that the Figure shows two consecutive events with no change to the optical or camera settings.

The microbunches observed on Fig. \ref{fig:streakimage} and the protons ahead of the laser pulse form the bunch core of Fig. \ref{fig:plasmaonoff}b). The defocused protons acquire a larger diverging angle along the bunch, as suggested by Fig. \ref{fig:streakimage}. On Fig. \ref{fig:plasmaonoff}b) they form a faint halo, below detection threshold, but are clearly visible on the halo camera image (Fig. 
d)). The effect of the transverse plasma wakefield on the proton bunch is clearly seen in the differences between Figs. \ref{fig:plasmaonoff}
c) and 
d) and is suggested by Figs. 
a) 
and 
b).

Figure \ref{fig:plasmaonoff}e) compares the vertical projections of the measurements shown in Figs. \ref{fig:plasmaonoff}a--d). Since we know the centroid position of the cores as well as the scale factor between the core and halo cameras from measurements without plasma and mask, we can combine the images from the core and halo to form one profile. Without plasma, there is a gap between the profiles, caused by the large difference in attenuation and the limited dynamic range of the cameras. We interpolate the profile between the distribution using a cubic 1D interpolation routine (blue dotted line).  

From the images and bunch centroid position, we determine the maximum radius of the self-modulated bunch distribution (as well as its uncertainty) with the contour method described in \cite{TurnerEAAC2017}. The resulting maximum radius is shown with green bars on Fig. \ref{fig:plasmaonoff}e). The halo is clearly observed on Fig. \ref{fig:plasmaonoff} b) and d) and extents to a radius of \unit[$r_{max} = (14.5\pm1.0)$]{mm}.

Figures \ref{fig:maxradius}b), 
d) and 
e) show that with the plasma, the peak intensity of the core image decreases as defocused particles leave the core for the halo. Integrating the areas under the blue and red curves, we find that the total number of counts on the image is conserved at the percent level, when normalized to the incoming charge. 

The figures also show that this increase in charge density at large radial positions is symmetric around the bunch center (as was the case for all measurements in this \textit{Letter}). This shows that the self-modulation process developed symmetrically along the plasma and suggests that the non-symmetric version of the process, known as the hose instability \cite{HOSING}, did not develop. This is consistent with numerical results \cite{HOSEVIERA,HOSESCHROEDER} that show that, although the two processes have a comparable growth rate, seeding of the symmetric self-modulation process can suppress the development of the asymmetric process.

The defocused protons at $r_{max}$ experienced the highest product of transverse wakefield amplitude and interaction time with the wakefields, and hence gained the largest transverse momentum. Figure \ref{fig:plasmaonoff}e) shows that for a plasma density of \unit[$7.7\times10^{14}$]{electrons/cm$^3$} defocused protons reach to a maximum radius of \unit[$r_{max} =$(14.5$\pm$1.0)]{mm}. The IS 2 is located \unit[$\sim$20]{m} downstream of the plasma entrance and the protons moving at the speed of light must acquire their transverse momentum before exciting the wakefields within a maximum time corresponding to a length of \unit[10]{m} of plasma. Their defocusing angle ($\theta$) must thus be between \unit[0.73]{mrad} (exit wakefield at \unit[z=0]{m}) and \unit[1.45]{mrad} (exit at \unit[z=10]{m}), which corresponds to a total transverse momentum between \unit[290 $\textrm{and}$ 580]{MeV/c}. 

From the defocusing angle $\theta$, we estimate the average transverse wakefield amplitude ($W_{\perp,av}$) that must have been driven by the self-modulating proton bunch. We assume that defocused protons experience  constant amplitude transverse wakefields, both along the plasma over a distance $L$ and radially until they exit the wakefields transversely:
\begin{equation}
W_{\perp,av} = \frac{\theta \cdot p_{\parallel}c}{q\cdot L}
\label{eq:angle}
\end{equation}
where $q$ is the proton charge and $p_{\parallel}$ the proton longitudinal momentum of \unit[400]{GeV/c}. The rms value of the emittance driven proton bunch divergence (\unit[$\theta_{\epsilon}=0.034$]{mrad}) is subtracted from $\theta$. We calculate the lowest limit of the wakefield amplitude $W_{\perp,av,min}$ from the measurements and Eq. \ref{eq:angle} by assuming that the protons experience their transverse momentum over the full plasma distance of \unit[$L=$10]{m}.

Since the bunch $n_b$ to plasma density $n_{pe}$ ratio is initially small, $n_b/n_{pe}=5\times10^{-3}$, we use linear plasma wakefield theory to calculate the transverse seed wakefields amplitude. We calculate it according to \cite{LINEARTHEORYEQ} for the case of a step density of the Gaussian proton bunch at the seed point one quarter $\sigma_z$ ahead of the center of the bunch. 

The maximum transverse initial seed wakefield amplitude for the experimental proton bunch and plasma parameters is \unit[$15$]{MV/m} (at their radial maximum $r\simeq\sigma_r$). Note that these initial transverse wakefields are only focusing and located close to the seed point.
The amplitude of the defocusing fields at $\xi=\sigma_z$ behind the bunch center yields only \unit[$\sim 6$]{MV/m}.

\begin{figure}[htb!]
\centering
		\includegraphics[width =\columnwidth]{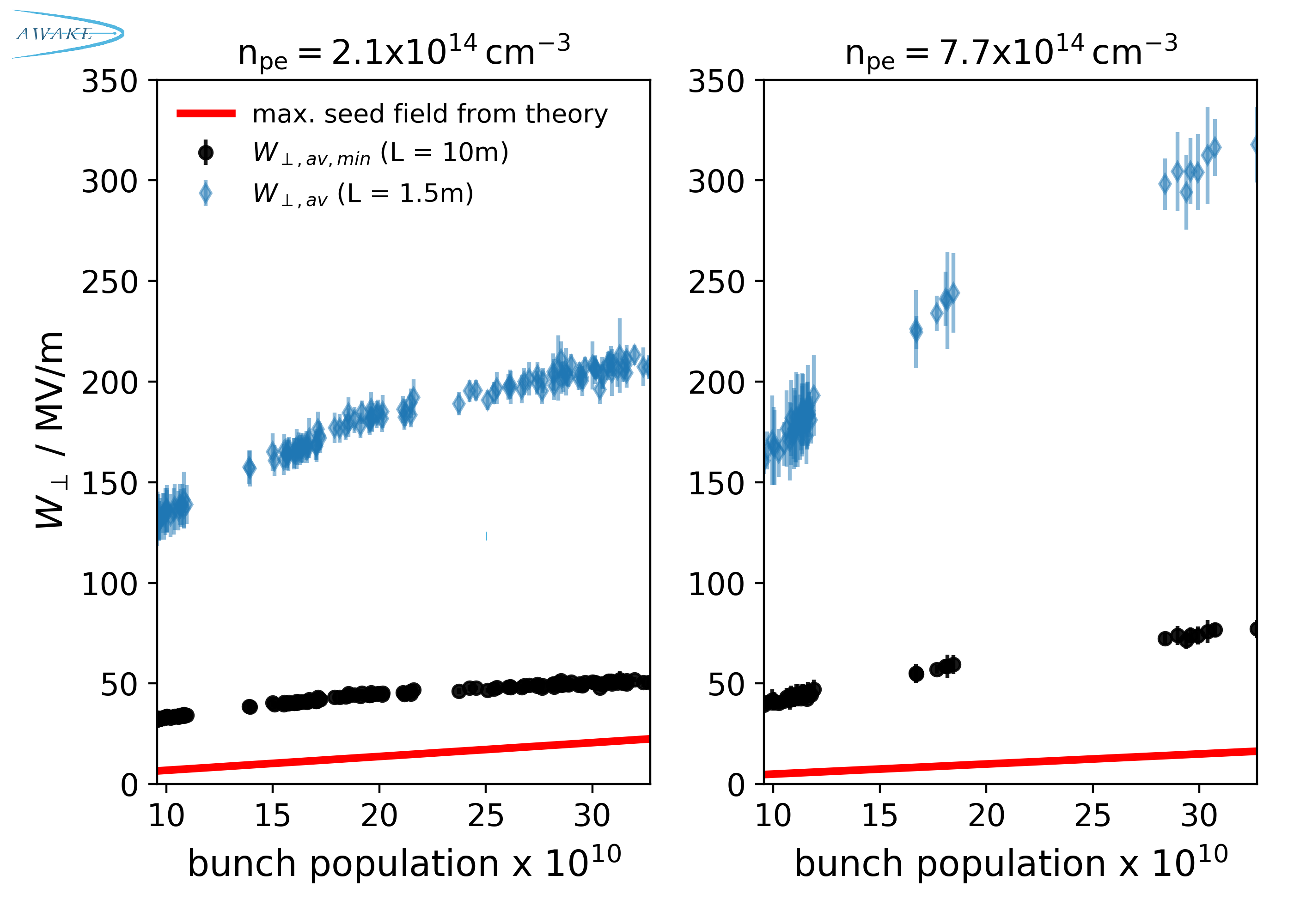}
		\caption{Transverse wakefield amplitude as a function of proton bunch population for two plasma electron densities \unit[$2.1\times10^{14}$]{electrons/cm$^3$} (left) and \unit[$7.7\times10^{14}$]{electrons/cm$^3$} (right). The red lines show the seed wakefield amplitude. The black dots show the lowest limit of the transverse wakefield amplitude $W_{\perp,av,min}$ assuming that the maximum defocused protons exit the wakefields after  \unit[L=10]{m}. The blue diamonds show the best estimate from simulations $W_{\perp,av}$ obtained with \unit[L=1.5]{m} and assuming that the protons exit at \unit[4]{m}.}     
\label{fig:growth}
\end{figure}

We obtained images similar to those of Fig. \ref{fig:plasmaonoff} as a function of the proton bunch population, with the laser pulse one-quarter $\sigma_z$ ahead of the bunch center and all other parameters kept constant. 

On Fig. \ref{fig:growth} we compare the calculated maximum initial seed field amplitude (red line) with the lowest limit of the wakefield amplitude ($W_{\perp,av,min}$, black dots, Eq. \ref{eq:angle}), as a function of the proton bunch charge for two different plasma electron densities. We observe that the minimum average wakefield amplitude increases with increasing proton bunch charge. This is as expected since both the initial wakefield amplitude and the growth rate increase with increasing bunch charge \cite{GROWTH}. 

Figure \ref{fig:growth} also shows that the lowest limit of the wakefield amplitude is larger than the initial seed amplitude for all measured proton charges, even under the very conservative assumptions used to estimate their values from the experimental data. In all cases presented here, $W_{\perp,av,min}$ is at least a factor 2.3  larger than the seed wakefield amplitude. At the highest plasma density and with the largest proton bunch population, they are a factor of 4.5 larger. It is clear evidence of the growth of the wakefields from their seed values along the plasma.

It is however also clear that the peak amplitude of the wakefield must be larger than these $W_{\perp,av,min}$ values since the wakefield amplitude: a) has a non-constant transverse dependency (zero on-axis and peak at $r=\sigma_r$) and b) grows along the plasma as stated above. 

From simulations results \cite{TURNERNAPAC2016}) we expect protons to radially exit the wakefields after \unit[$\approx$4]{m}, much earlier than the \unit[10]{m} used above. Simulations and estimates show that the strongly defocused protons gain most of their transverse momentum over \unit[$\sim$0.5-1.5]{m}, since the wakefields amplitude at the beginning of the plasma is small. The blue diamonds in Fig. \ref{fig:growth} show the average transverse wakefield amplitude from the measurements ($W_{\perp,av}$), assuming protons interact with the plasma over a distance of \unit[1.5]{m} and exit the plasma at \unit[4]{m}. In this case the average wakefield amplitude reaches hundreds of MV/m.


Simultaneously to the symmetric defocusing of the protons on IS 1 and 2, we observe the formation of microbunches on the streak camera diagnostic \cite{karl,STREAK}, see Figure \ref{fig:streakimage}. This is proof for successful radial self-modulation over the \unit[10]{m} of plasma.



The experimental results presented here show that the time structure of the relativistic proton bunch exiting the \unit[10]{m}-long plasma is due to periodic defocusing along the bunch. They show that defocusing increases along the bunch and along the plasma. The transverse wakefields causing the defocusing exceed the seed amplitude value (\unit[<15]{MV/m}) and reach over \unit[300]{MV/m}. The defocusing is symmetric around the bunch propagation axis. These results therefore show that the seeded self-modulation of the proton bunch occurred along the long bunch and suggest that its non-axis-symmetric counterpart, the hose instability did not develop. Together with the excitation of the transverse wakefields causing the effects reported here, come longitudinal wakefields. These components 
have been used to accelerate externally, low energy injected electrons (\unit[10-20]{MeV}) to multi-GeV energy levels~\cite{ILOVEEA} 
and possibly to hundreds of GeVs or TeVs in the future and for high-energy physics applications \cite{PROTONDRIVEN,VHEep}.

\section*{Acknowledgements}
This work was supported in parts by the Siberian Branch of the Russian Academy of Science (project No. 0305-2017-0021), a Leverhulme Trust Research Project Grant RPG-2017-143 and by STFC (AWAKE-UK, Cockroft Institute core and UCL consolidated grants), United Kingdom; a Deutsche Forschungsgemeinschaft project grant PU 213-6/1 ``Three-dimensional quasi-static simulations of beam self-modulation for plasma wakefield acceleration''; the National Research Foundation of Korea (Nos.\ NRF-2015R1D1A1A01061074 and NRF-2016R1A5A1013277); the Portuguese FCT---Foundation for Science and Technology, through grants CERN/FIS-TEC/0032/2017, PTDC-FIS-PLA-2940-2014, UID/FIS/50010/2013 and SFRH/IF/01635/2015; NSERC and CNRC for TRIUMF's contribution; and the Research Council of Norway. M. Wing acknowledges the support of the Alexander von Humboldt Stiftung and DESY, Hamburg. The AWAKE collaboration acknowledge the SPS team for their excellent proton delivery.

\end{document}